# Revised Group Sunspot Number for 1640, 1652, and 1741

J. M. Vaquero[1,2] • R. M. Trigo[2,3]

[1]Departamento de Física, Universidad de Extremadura, Avda. Santa Teresa de Jornet, 38, 06800 Mérida (Badajoz), Spain (e-mail: jvaquero@unex.es)

[2]CGUL-IDL, Universidade de Lisboa, Lisbon, Portugal

[3]Departamento de Eng. Civil da Universidade Lusófona, Lisbon, Portugal

**Abstract**

Some studies have shown that our knowledge on solar activity in the years 1640, 1652, and 1741 can be improved. In this contribution, we revise the annual group sunspot numbers for these years from original observations. For the years 1640, 1652 and 1741, we have obtained the corrected values 15.2, 1.8, and 27.3, respectively (instead of the original values 15.0, 4.0, and 57.7).

**1. Introduction**

The sunspot number is the longest instrumental record of solar activity available covering the last four centuries. Since 1980, the Royal Observatory of Belgium produces the International Relative Sunspot Number (Clette et al., 2007). Hoyt and Schatten (1998), hereafter HS98, derived a new reconstruction of solar activity since 1610 from historical sunspot observations (Vaquero, 2007; Usoskin, 2013) called Group Sunspot Number (hereafter GSN). Despite the tremendous amount of work required by Hoyt and Schatten to compile the GSN, several recent studies have underpinned the existence of some inconsistencies and a few potential erroneous values. Thus, in the last years, at least three important changes have been proposed to improve the reliability of the GSN series, namely in relation to the following periods, (i) the onset of Maunder Minimum (Vaquero et al., 2011) around 1640, (ii) the solar cycle -1 (Vaquero, Gallego and Trigo, 2007), and (iii) the "lost" solar cycle (Usoskin et al., 2009; Zolotova and Ponyavin, 2011). It is important to stress that GSN series are widely used for long-term solar variability studies and, therefore, this changes have direct consequences in other series (as Total Solar Irradiance) derived from GSN series. Moreover GSN series are often used in climate modelling studies that attempt to reconstruct global or regional climate variability and trends at the centennial and millennial scales (IPCC, 2007).

It is within this context that the authors have undertaken previous efforts to detect some of the most likely serious problems with the GSN series (Vaquero, Gallego and Trigo, 2007; Vaquero et al., 2011) including the application of a simple empirical rule to detect anomalous values in the GSN annual series (Vaquero, Trigo and Gallego, 2012). In particular, the years 1652 (during the Maunder Minimum) and 1741 (during the solar cycle -1) were flagged as problematic years.

The aim of this short contribution is to provide a revised estimation of GSN values for three of these "problematic years" (1640, 1652, and 1741) improving significantly the reliability of the GSN values for those years. In this regard, we are confident that this work will help solving some problems of the historical part of this valuable solar index.

**2. Sunspots during 1640**

The year 1640 is very poorly covered by solar observations. While 17 and 42 records are preserved for the years 1637 and 1642, only two records are preserved from 1640. This is particularly problematic as an erroneous GSN value for this year can lead to incorrect assumptions in our understanding of how the Maunder minimum started. Therefore, we must emphasise that in the HS98 database there are only two observations for this year made by Scheiner in early summer (21-22 June).

We add two new recovered records by Polish astronomer Albert Strażyc from Krakow (Poland). The observations were made in late summer, the days 5 (at 03:00 pm) and 6 (at 11:00 am) September. These observations are described in the leaflet entitled *Quaestio astronomica* (f. A3 bis verso). Dobrzycki (1999) has studied in detail this text proposing that the observers were probably Stanisław Pudłowski and Albert Strażyc. A sketch containing the positions of the sunspot group was also published (see Dobrzycki, 1999, p. 126). Note that we double the number of available observations (from 2 to 4) when we incorporate this Polish record into the GSN database. However, the annual value does not change (except the rounding) because we have now four records of one sunspot group instead two records of one group.

## 3. Sunspots during 1652

The year 1652 is one of the problematic years detected by Vaquero, Trigo and Gallego (2012) in the HS98 database. HS98 listed only three sunspot observers for this year, including: 1) Hevelius (who observed some sunspots in April and no sunspots in some days of October, November and December), 2) Petitus (who observed no spots on 8 April), 3) an unknown observer (who observed no spot from 14 November to 31 December).

The source used by HS98 to obtain the sunspot observations by Hevelius was the well-known *Mittheilungen über die Sonnenflecken*, published by R. Wolf. Note that Hoyt and Schatten (1995) revised other sunspot records by Hevelius. The available information on sunspot in 1652 was first published by Hevelius (1652). Later, Wolf (1856) published an abstract of the original source. The exact descripction made by Wolf is:
"75) Illustribus Viris, Petro Gassendo et Ismaeli Bullialdo, Johannis Hevelius.
Acht, »Gedani 1652 die 10. Julii st. n.« datirte Folioseiten über die Sonnenfinsterniss vom 8. April 1652. Er erzählt, dass er am 1. April 5 Flecken, am 3. noch 2 gesehen habe, die aber am 6. in Fackeln degenerirt seien, so dass man am 7. und 8. April gar nichts in der Sonne gesehen habe." (Wolf, 1856, p. 151)

We have located the original text (in Latin): "[...] *quanquam die 1. Aprilis, horâ 11. 45'. in disco Solis quinq; visae fuerint maculae: duae quidem debilissimaes non procul à limbo orientali, dilutioribus concomitantibus faculis umbrisq; tres autem fatis densae, circa centrum, in latitudine Boreali. Ex quibus posterioribus die 3. Aprilis tantúm duae conspectae, quae die sextâ in faculas penitus degeneravére; reliquae verô duae debiliores, die 7. omnino etiam sunt exstinctae.*" (Hevelius, [1652]). A modern translation could be: "[...] However, on 1st April at 11 hours and 45 minutes five spots were seen on the solar disk, two weak spots near the eastern limb between faculae and umbrae, and three dense spots near the centre, in north latitude. Later, on 3 April , only two spots were seen, which completely degenerated into faculae on day 6. On day 7, the remains of the two weakest spots also disappeared."

According to the original text, on the day 1 April there were only two groups (containing five spots) on the Sun (instead of five groups). The first group, composed by two weak spots, was located close to the eastern limb and the second group, composed by three spots, was located near the centre of the solar disc (in the northern hemisphere). It is clear that the text does not provide a detailed description of sunspots. However, we are confident to suppose that the two spots that are mentioned on 3 April correspond to the first group detected earlier. Therefore, we have two groups for 1 April and only one group remains for 3 April. No spots were observed in 6 and 7 April.

Following this simple assumption we can assign a monthly value of sunspot groups equal to 0.6 using four observations by Hevelius and one by Petitus. The annual value is equal to 0.15 groups according to the monthly values 0.60 for April and 0.00 for October, November and December. Therefore, the annual value for GSN is 1.8 using the definition by HS98 (who assigned 4.0 to this annual value).

**4. Sunspots during 1741**

The solar cycle number -1 (1733-1744 approximately) is very poorly covered by observations according to HS98. In fact, this solar cycle presents an unusual shape with three peaks in the years 1736, 1738, and 1741 (see Figure 1, red line). According to HS98 the year of maximum solar activity corresponds to 1741, *i.e.* very delayed in respect to the usual shape of a common solar cycle. A statistical method (Usoskin, Mursula, and Kovaltsov, 2003) devised to obtain a better estimate of the annual value of the NSG when there is a small number of observations available clearly showed that the value for 1741 was significantly overestimated. Vaquero, Gallego and Trigo (2007) improved the initial unusual shape of this solar cycle using information about solar activity for a four-years period (1736-9) published in three journals of that epoch: *Philosophical Transactions*, *Histoire de l'Académie Royale des Sciences*, and *Nova Acta Eruditorum*. However, despite these improvements between 1736 and 1739, the unusual peak of the year 1741 remained as the maximum of this solar cycle.

Note that there is a large gap in sunspot observations during four consecutive years (1744-7). This wide gap has important consequences for the reconstruction of solar activity from historical sunspot observations because it prevents the clear determination of the GSN evolution between cycles -1 and 0 but also undermines the comparisons between different observers in order to obtain calibration factors.

Obviously, very few records of sunspot observation for 1741 are preserved in archives and libraries. HS98 database only contains two observers for this year. The first observer is Musano who observed no spots for 16 and 24-28 December and one group for 17-23 and 29-31 December (monthly value of GSN is equal to 9.1). The second observer is John Winthrop who observed seven sunspot groups for 10 January according to HS98.

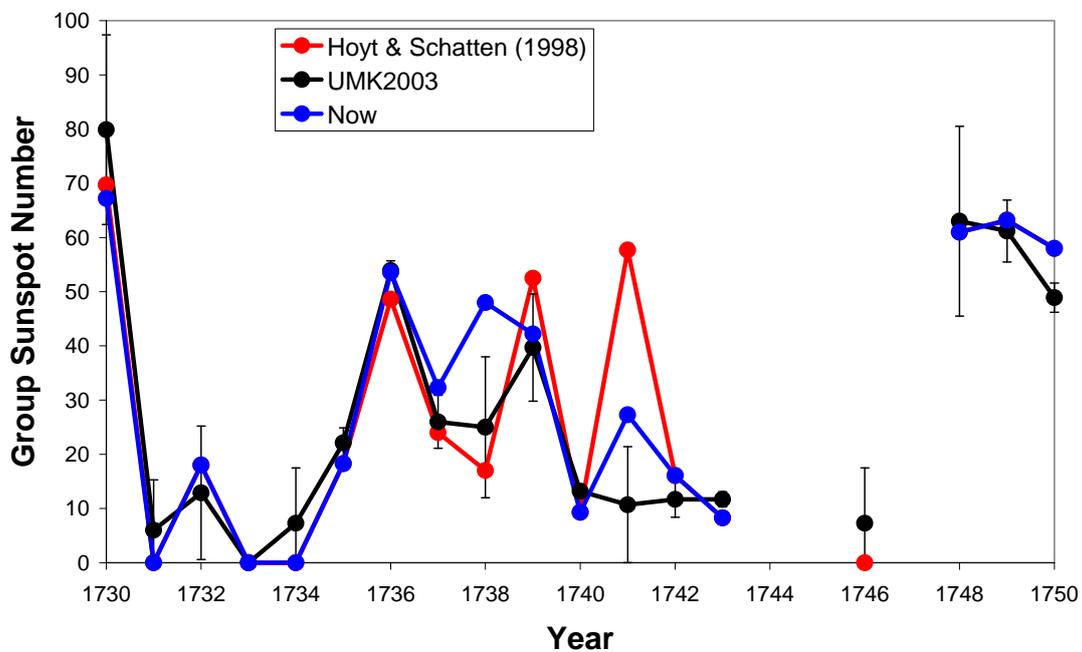

Figure 1. Annual values of GSN in Hoyt and Schatten (1998) (red), according to a statistical method by Usoskin, Mursula and Kovaltsov (2003) (black) and using the cumulative corrections proposed by the authors in Vaquero, Gallego and Sánchez-Bajo (2007), Vaquero, Gallego and Trigo (2007), and this work (blue).

We have located the original manuscript of this report by Winthrop in Harvard University Archives (Papers of John and Hannah Winthrop, HUM 9, Box 4, Volume 2). The original report state: "10. noon. the great number of spots in the Sun I ever saw. One I discover with my naked eye (with only a colored glass to save it) with through telescope appeared to be a cluster of spots exceeding black & incomparable on all sides with a nebula: & besides you, they were 5 or 6 in other parts of the Sun [...]"

We can note that it is not possible to establish an exact count of groups and spots. If we suppose that the "5 or 6" spots are clustered in two groups, we have three groups for that day. It corresponds to a daily GSN value equal to 45.5 assuming that the calibration constant of Winthrop is 1.255 (HS98). Therefore, the annual GSN value for this year is 27.3 obtaining a result more in line with the one obtained by Vaquero, Trigo and Gallego (2012) using their simple method (Figure 1).

**5. Conclusions**

In this work, we have revised some historical sunspot observations associated with problematic annual values of GSN. Despite the small number of observations, the results of the new estimates of annual sunspot number are of interest because we have worked with years in which there are very few observations.

Table 1. Annual GSN values for the years 1640, 1652 and 1741 from HS98 (second column) and this work (third column).

| Year | HS98 | This work |
|------|------|-----------|
| 1640 | 15.0 | 15.2 |
| 1652 | 4.0  | 1.8 |
| 1741 | 57.7 | 27.3 |

We have doubled the number of known observations made in the year 1640 but only four records were preserved and the GSN for that year remained similar to that put forward by HS98. For the years 1652 and 1741, we have revised the sunspot group numbers values consulting the original historical sources. Using these new values, we have computed new annual GSN (see Table 1) that improves our knowledge of solar activity for these critical years.

**Acknowledgements**

Authors thank the Harvard University Library that provided a copy of the observation of Winthrop. J.M. Vaquero has benefited from the impetus and participation in the Sunspot Number Workshops (http://ssnworkshop.wikia.com/wiki/Home). Support from the Junta de Extremadura (Research Group Grant No. GR10131) and Ministerio de Economía y Competitividad of the Spanish Government (AYA2011-25945) is gratefully acknowledged. Ricardo M. Trigo was supported by Portuguese Science Foundation (FCT) through project QSECA (PTDC/AAG-GLO/4155/2012).

**References**

Clette, F., Berghmans, D., Vanlommel, P., van der Linden, R.A.M., Koeckelenbergh, A., Wauters, L.: 2007, *Adv. Space Res.* **40**, 919.

Dobrzycki, J.: 1999, *Journal for the History of Astromomy* **30**, 121.

Hevelius, J.: [1652], *Illustribus Viris, Petro Gassendo, & Ismaeli Bullialdo, Philosophis ac Mathematicis nostri seculi summis, amicis suis officiose honorandis* [Danzig].

Hoyt, D. V., Schatten, K. H.: 1995, *Solar Phys.* **160**, 371.

Hoyt, D. V., Schatten, K. H.: 1998, *Solar Phys.* **179**, 189.

IPCC: 2007, *Climate Change 2007: The Physical Science Basis. Contribution of Working Group I to the Fourth Assessment Report of the Intergovernmental Panel on Climate Change*, Cambridge University Press, Cambridge, United Kingdom and New York.

Usoskin, I.G.: 2013, *Living. Rev. Solar Phys.* **10**, 1.

Usoskin, I.G., Mursula, K., Kovaltsov, G.A.: 2003, *Solar Phys.* **218**, 295.

Usoskin, I.G., Mursula, K., Arlt, R., Kovaltsov, G.A.: 2009, *Astrophys. J. Lett.* **700**, L154.

Vaquero, J.M.: 2007, *Adv. Spa. Res.* **40**, 929.

Vaquero, J.M., Gallego, M.C., Sánchez-Bajo, F.: 2007, *The Observatory* **127**, 221.

Vaquero, J.M., Gallego, M.C., Trigo, R.M.: 2007, *Adv. Space Res.* **40**, 1895.

Vaquero, J.M., Trigo, R.M., Gallego, M.C.: 2012, *Solar Phys*. **277**, 389.

Vaquero, J.M., Gallego, M.C., Usoskin, I.G., Kovaltsov, G.A.: 2011, *Astrophys. J. Lett.* **731**, L24.

Wolf, R.: 1856, *Mittheilungen über die Sonnenflecken* **6**, 151.

Zolotova, N.V., Ponyavin, D.I.: 2011, *Astrophys. J.* **736**, 115.